\documentclass[twocolumn,aps,prx,showpacs,floatfix]{revtex4-1}
\usepackage{amsmath}
\usepackage{amsfonts}
\usepackage{amssymb}
\usepackage{graphicx}
\usepackage{bm}
\usepackage{bbold}
\usepackage{float}
\usepackage{hyperref}
\usepackage{color}
\usepackage{tabularx}

\begin{document}
\title{Tunable Majorana corner states in a two-dimensional second-order topological superconductor induced by magnetic fields}
\author{Xiaoyu Zhu}
\affiliation{School of Science, Xi'an Jiaotong University, Xi'an, Shaanxi 710049, China}

\date{\today}

\begin{abstract}
A two-dimensional second-order topological superconductor exhibits a finite gap in both bulk and edges, with the nontrivial topology manifesting itself through Majorana zero modes localized at the corners, \textit{i.e.}, Majorana corner states. We investigate a time-reversal-invariant topological superconductor in two dimension and demonstrate that an in-plane magnetic field could transform it into a second-order topological superconductor. A detailed analysis reveals that the magnetic field gives rise to mass terms which take distinct values among the edges, and Majorana corner states naturally emerge at the intersection of two adjacent edges with opposite masses. With the rotation of the magnetic field, Majorana corner states localized around the boundary may hop from one corner to a neighboring one and eventually make a full circle around the system when the field rotates by $2\pi$. In the end we briefly discuss physical realizations of this system.
\end{abstract}


\maketitle

\section{Introduction}

Majorana zero modes (MZMs), being mid-gap bound states, are defining features of topological superconductors (TSCs). Just like Majorana fermions\cite{Majorana1937}, a MZM is also the anti-particle of itself, usually denoted by a self-adjoint operator $\gamma=\gamma^\dagger$\cite{Wilczek2009}. The last decade has witnessed a rapid development in the pursuit of MZMs\cite{Read2000,Kitaev2001,Fu2008,Lutchyn2010,Oreg2010,Alicea2012,Elliott2015,Beenakker2013,Stanescu2013,Aguado2017}, with signatures being reported recently in various systems, such as in nanowire (atomic chain)/superconductor\cite{Mourik2012,Das2012,Deng2012,Rokhinson2012,Finck2013,Perge2014,Deng2016} or topological insulator/superconductor heterostructure\cite{Xu2015,Sun2016}, to name a few. Essentially, these systems could realize TSCs under specific circumstances, and MZMs in general emerge at domain walls, for instance, boundaries or vortices, across which the bulk gap of a TSC reverses sign, hence signifying a change of topology.

In contrast to traditional 2D (3D) topological systems where protected gapless modes usually occur on edges (surfaces), very recently it was proposed that topologically nontrivial modes could also emerge at corners (hinges) of certain 2D (3D) systems, coined higher-order topological insulators or superconductors\cite{Sitte2012,Teo2013,Zhang2013a,Benalcazar2014,Benalcazar2017,Benalcazar2017prb,Schindler2017,Langbehn2017,Song2017,Imhof2017,Peterson2017,Kunst2017,Ezawa2018,Serra-Garcia2018}. A second-order TSC in 2D, according to the definition, is characterized by Majorana corner states (MCSs), \textit{i.e.}, MZMs bound at corners, where two topologically distinct edges intersect and give rise to a domain wall resembling that in traditional TSCs. Creating such a domain wall at the intersection of neighboring edges is crucial to the realizations of second-order TSCs. To achieve this, one may enforce certain (spatial) symmetries in a traditional TSC at start, as in Ref.[\onlinecite{Langbehn2017}], where two adjacent edges related by reflection symmetry exhibit gaps of opposite signs, and a symmetry-breaking perturbation weak enough could not immediately eliminate the sign differences and hence the domain wall survives. Alternatively, one could start from a domain wall separating two gapless systems with distinct topology, and apply an external field to gap them out, as was investigated in Ref.[\onlinecite{Volovik2010}] on the surface of $^3\text{He-B}$, where magnetic field acting on two gapless domains that are characterized by opposite Ising variables could introduce masses of reverse signs, and therefore a chiral MZM forms at the domain wall.

Up to now, research on second-order TSCs are still at initial stage, with only a few models being put forward to support MCSs\cite{Teo2013,Benalcazar2014,Langbehn2017}, and it remains unknown whether these proposals would eventually lead to experimental realizations. In this context, it would be worthwhile to look for other simple as well as physically relevant systems that may support MCSs. In this work, we start from a $p\pm ip$ superconductor, the minimal model of 2D time-reversal invariant TSCs which belong to DIII class\cite{Schnyder2008,Chiu2016}, and propose that this simple system could support MZMs at its corners when an in-plane magnetic field is applied. We demonstrate that a uniform magnetic field could gap out the edges, whereas edge gaps may reverse signs across certain corners, thus accommodating MZMs inside. By mapping all the four edges to a 1D system, we develop an effective edge theory in which a uniform Zeeman field is projected to a spatially varying mass field, and a single MCS emerges naturally when a kink forms at the intersection of two neighboring edges. In the end, we briefly discuss a more realistic model, Rashba semiconductor/nodeless iron-based superconductor heterostructure in two dimension introduced in Ref.[\onlinecite{Zhang2013b}], and demonstrate that an in-plane magnetic field could also give rise to MZMs at the corners of this system. Our research may stimulate future searches for MCSs on physical systems that could potentially realize time-reversal-invariant TSCs.

\section{Model}

A $p\pm ip$ superconductor in 2D is characterized by cooper pairing in $p+ip$ form for spin-up (down) electrons and $p-ip$ for spin-down (up) electrons\cite{Qi2009}. 
Consider a $2L \times 2L$ square lattice for such a system when subject to an in-plane magnetic field, with tight binding Hamiltonian given by
\begin{eqnarray}
&&H = -t\sum\limits_{\langle\bm{rr}'\rangle\alpha} c^\dagger_{\bm r\alpha} c_{\bm r'\alpha} + \sum\limits_{\bm r \alpha\alpha'} c^\dagger_{\bm r\alpha}(\mu\sigma_0+\bm V\cdot\bm\sigma)_{\alpha\alpha'}c_{\bm r\alpha'}\nonumber\\
 &&+ \frac{\Delta}{2} \sum\limits_{\bm r\alpha} e^{is_\alpha\phi}(c^\dagger_{\bm r\alpha}c^\dagger_{\bm r + \bm{\hat{e}}_y,\alpha} - is_\alpha c^\dagger_{\bm r\alpha}c^\dagger_{\bm r+\bm{\hat{e}}_x,\alpha} ) + \text{H.c.},\label{eq:Hamiltonian}
\end{eqnarray}
where $t$ terms include only nearest-neighboring hopping, $\mu$ denotes chemical potential, $\bm V = (V_x,V_y,0)$ represents the Zeeman field induced by an external magnetic field in the plane, $\bm{\hat{e}}_x$ and $\bm{\hat{e}}_y$ denote unit vectors along $x$ and $y$ directions, $\bm \sigma$ represent Pauli matrices, $\sigma_0$ stands for identity matrix, and $\alpha$ refers to spin indices with $s_\uparrow = 1$ and $s_\downarrow=-1$. In this model, spin-up and spin-down electrons condense into cooper pairs of $p-ip$ and $p+ip$ form separately, with pairing strength $\Delta$ being the same for both and the global superconducting phase $\phi$ taking opposite signs as is guaranteed by time reversal symmetry (TRS). 

\begin{figure}[t]
\includegraphics[scale=0.25]{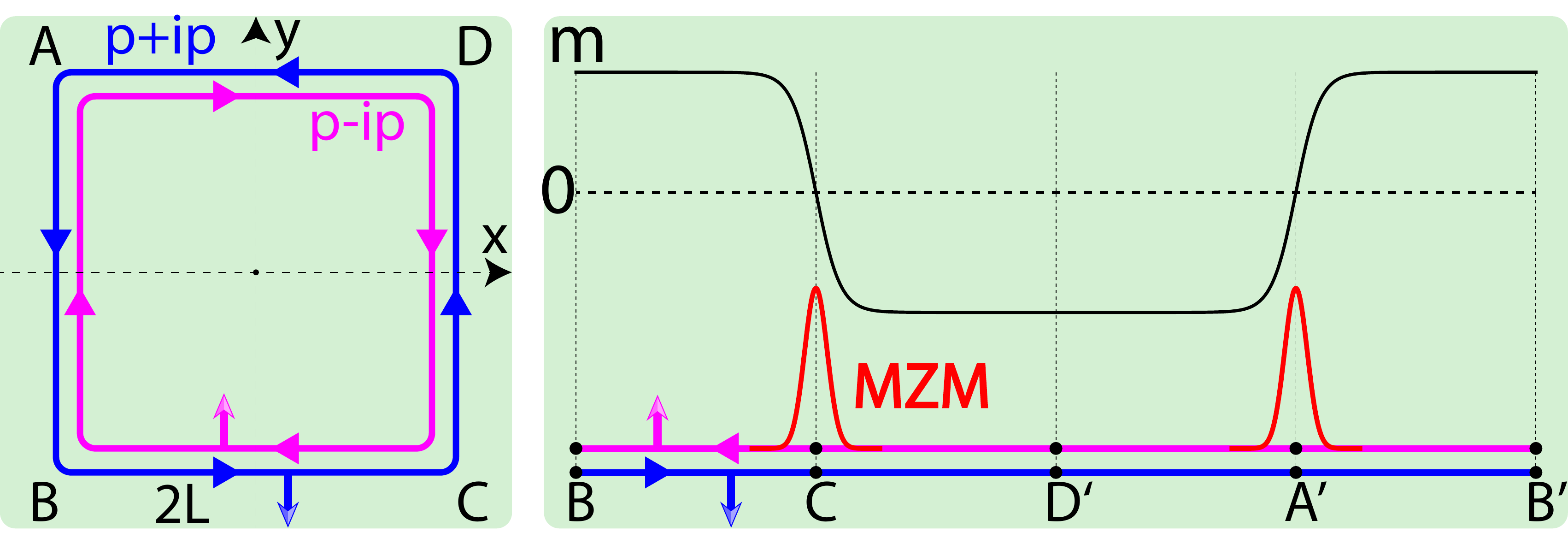}
\caption{Left panel. Schematic plot of a 2D $p\pm ip$ superconductor, which features counter-propagating edge modes, with spin-up modes denoted in magenta and spin-down modes in blue. Right panel. An effective description of the four edges with a 1D theory, where the letters $D'$, $A'$ and $B'$ correspond to their counterparts $D$, $A$ and $B$ in the left panel. In presence of an in-plane Zeeman field, the mass gap $m$ induced varies across edges and MZMs emerge at corners where $m$ reverses sign and forms a kink. }\label{fig:fig1}
\end{figure}

Imposing periodic boundary conditions (PBCs) in both directions, we could obtain the momentum representation of lattice Hamiltonian (\ref{eq:Hamiltonian}), which takes the form
\begin{eqnarray}
&&H = \frac{1}{2}\sum\limits_{\bm k} \Psi_{\bm k}^\dagger \mathcal{H}(\bm k)\Psi_{\bm k} \label{eq:Hamiltonian_k}\\
&&\mathcal{H}(\bm k) = \epsilon(\bm k)\tau_z - \Delta\tau_x(\sin k_x\sigma_x-\sin k_y\sigma_y) + \tilde{\bm V}\cdot\bm\sigma,\nonumber
\end{eqnarray}
where $\tau_i$ and $\sigma_i$ $(i = x, y, z)$ are Pauli matrices acting in particle-hole and spin space respectively, the Nambu spinor $\Psi_{\bm k}=e^{-i\frac{\phi}{2}\sigma_z}\{c_{\bm k\uparrow},c_{\bm k\downarrow},c^\dagger_{\bm {-k}\downarrow},-c^\dagger_{\bm {-k}\uparrow}\}^T$ (note the prefactor is added to eliminate the phase $\phi$ in superconducting terms), kinetic energy $\epsilon(\bm k) = \mu-2t(\cos k_x + \cos k_y)$, and the effective Zeeman field $\tilde{\bm V} = (V\cos(\theta+\phi),V\sin(\theta+\phi),0)$, with $\theta = \arg(V_x+iV_y)$. 

In absence of Zeeman fields, the system preserves time reversal symmetry $\mathcal T = i\sigma_y\mathcal K$ ($\mathcal K$ denotes conjugation operator), particle-hole symmetry $\mathcal P = \tau_y\sigma_y\mathcal K$, and in addition $C_4$ rotation symmetry given by
\begin{equation}
\mathcal U_R \mathcal H(\mathcal R^{-1}\bm k)\mathcal U_R^{-1} = \mathcal H(\bm k),\label{eq:c4rotaion}
\end{equation}
with $\mathcal U_R = e^{i\frac{\pi}{4}\sigma_z}$, and $\mathcal R (k_x, k_y) = (-k_y, k_x)$. Note that $\mathcal U_R$ now describes a clockwise rotation by $\frac{\pi}{2}$ about $\sigma_z$ axis, differing from the one usually defined for a $C_4$ rotation, say in Ref. \onlinecite{Schindler2017}, by a unitary transformation in spin space. Turning on a finite Zeeman field breaks TRS and $C_4$ rotation symmetry, but one can in this circumstance define an inversion symmetry $\mathcal I$ up to a gauge transformation, which reads
\begin{equation}
\mathcal U_I\mathcal H(\mathcal I^{-1}\bm k)\mathcal U_I^{-1} = \mathcal H(\bm k),\label{eq:inversion}
\end{equation}
with $\mathcal U_I  = \tau_z$ and $\mathcal I\bm k = -\bm k$. The bulk spectrum is supposed to preserve this inversion symmetry, and takes the following form
\begin{equation}
E(\bm k) = \pm\sqrt{(\sqrt{\epsilon^2(\bm k)+\Delta_-^2(\bm k)}\pm V)^2+\Delta^2_+(\bm k)},\label{eq:spectrum}
\end{equation}
with $\Delta_-(\bm k)=\Delta[\cos(\theta+\phi)\sin k_x-\sin(\theta+\phi)\sin k_y]$, and $\Delta_+(\bm k) = \Delta[\sin(\theta+\phi)\sin k_x+\cos(\theta+\phi)\sin k_y]$. The model in absence of Zeeman fields admits gapless modes only when $\mu$ is fine tuned to $0$ or $\pm 4t$. Away from these phase transition points when $\mu\in(-4t,0)\cup(0,4t)$, the system is fully gapped and resides in topologically nontrivial phases, in the sense that it cannot be smoothly connected to the vacuum and hence gapless modes would emerge at each edge, as illustrated in the left panel of Fig. \ref{fig:fig1}. Zeeman fields may gap out the edge modes and meanwhile reduce the gap size at Brillouin zone center $\Gamma (0, 0)$, corners $M (\pi, \pi)$ and centers of edges $X (0/\pi,\pi/0)$, as can be seen in Eq.(\ref{eq:spectrum}). A strong field, however, would drive the system into a nodal superconductor, with nodal points appearing in pairs due to inversion symmetry. In this work, we mainly work in the regime where the Zeeman field is weak enough such that the bulk gap in topologically nontrivial phases is always finite. This way one could focus on the edges, which we shall investigate in the following.

\section{Edge Hamiltonian}\label{sec:Edge}

As a first step, let us turn off the Zeeman field for the moment and write down the wave functions of MZMs on each edge following Ref. [\onlinecite{Read2000}], which are given by
\begin{equation}
\Psi^{\text{Edge}}_{\alpha}(r) = \mathcal A\psi^{\text{Edge}}_{\alpha}e^{\frac{\delta}{\Delta}\int_{\delta L}^r dr' [4t-\mu(r')]},\label{eq:wavefunction_edge}
\end{equation}
where the spinors $\psi^{\text{Edge}}_{\uparrow} =\{-e^{i\varphi_n},0,0,e^{-i\varphi_n}\}^T$, $\psi^{\text{Edge}}_{\downarrow} = \{0,e^{-i\varphi_n},e^{i\varphi_n},0\}^T$, with $\varphi_n = \frac{n}{4}\pi$, $\mathcal A$ is the normalization constant, and a domain wall is imposed around the edge considered, with $\mu(r)$ being assumed to vary slowly and less (greater) than $4t$ inside (outside) the system. Without loss of generality, $t$ and $\Delta$ are assumed to be positive. Variables $n$, $\delta$, $r$ appearing in Eq.(\ref{eq:wavefunction_edge}) take different values depending on the edge considered and are listed in the following table.
\begin{center}
\begin{tabularx}{0.48\textwidth}{ >{\setlength\hsize{0.48\hsize}\centering}X|>{\setlength\hsize{0.48\hsize}\centering}X|>{\setlength\hsize{0.48\hsize}\centering}X|>{\setlength\hsize{0.48\hsize}\centering}X|>{\setlength\hsize{0.48\hsize}\centering}X }
\hline
Edge         & AB & BC & CD & DA \tabularnewline \hline
$n$           & 1 & 2 & 3 & 4  \tabularnewline \hline
$r$            & $x$ & $y$ & $x$ & $y$ \tabularnewline \hline
$\delta$    & $-$ & $-$ & $+$ & $+$ \tabularnewline \hline
$\nu$  & $+$ & $-$ & $-$ & $+$ \tabularnewline \hline
\end{tabularx}
\end{center}

We note that, in deriving the wave functions of MZMs on a given edge, PBC is imposed along the direction of the edge considered, and $\mu(r)$ is assumed to be uniform along the same direction and to be close to $4t$ as well, in which case we could linearize the bulk Hamiltonian around $\Gamma$ point. Similarly, one may consider the case when $\mu$ is close to the other two phase transition points, and in those circumstances linearize the Hamiltonian around $M$ or $X$ points, which would lead to the same results except that the wave functions for edge modes may acquire additional oscillating terms like $e^{i\pi r}$. Clearly MZMs in Eq.(\ref{eq:wavefunction_edge}) always come in pairs, one for each spin. Upon projecting the bulk Hamiltonian (\ref{eq:Hamiltonian_k}) onto each edge space spanned by the basis $\{\Psi^{\text{Edge}}_{\uparrow}(r),\Psi^{\text{Edge}}_{\downarrow}(r)\}^T$ for a given edge, one could then obtain corresponding edge Hamiltonian, which reads
\begin{equation}
\mathcal H^{\text{Edge}}(k_{\bar r}) = \nu\Delta k_{\bar r} \eta_z^{\text{Edge}}- V\sin(\theta+\phi+2\varphi_n)\eta_y^{\text{Edge}}, \label{eq:Hamiltonian_edge}
\end{equation}
where $\eta_i^{\text{Edge}}$ are Pauli matrices acting in the edge space of a given edge, $\nu$ for each edge is listed in the table above, $\bar r = x$ if $r = y$ and vice versa. In Eq.(\ref{eq:Hamiltonian_edge}), the kinetic term proportional to $\Delta $ exactly describes two Majorana modes propagating in opposite directions, whereas the $\eta_y^{\text{Edge}}$ (mass) term which originates from the Zeeman field couples the two modes, thus opening a finite edge gap. As we have mentioned earlier, MCSs could form when two adjacent edges acquire masses of opposite signs. Although the edge gap size differs among the four edges, as can be seen in Eq.(\ref{eq:Hamiltonian_edge}), we cannot however infer the sign of each edge gap from it. Note that one can always reverse the sign of $\eta_y^{\text{Edge}}$ term in Eq.(\ref{eq:Hamiltonian_edge}) by choosing a different gauge for $\psi_\alpha^{\text{Edge}}$ in Eq.(\ref{eq:wavefunction_edge}), for instance, multiplying $\psi_\uparrow^{\text{Edge}}$ by a prefactor $e^{i\pi/2}$ and $\psi_\downarrow^{\text{Edge}}$ by $e^{-i\pi/2}$ at the same time. This way, the relative signs of gaps for adjacent edges would depend on the gauges chosen for each edge, which is clearly not valid. In the following, we shall identify the relative signs of the mass terms, and we will show that wave functions of MZMs among the four edges are actually related to one another through certain transformations.  


\section{Majorana corner states}

\begin{figure}[t]
\includegraphics[scale=0.6]{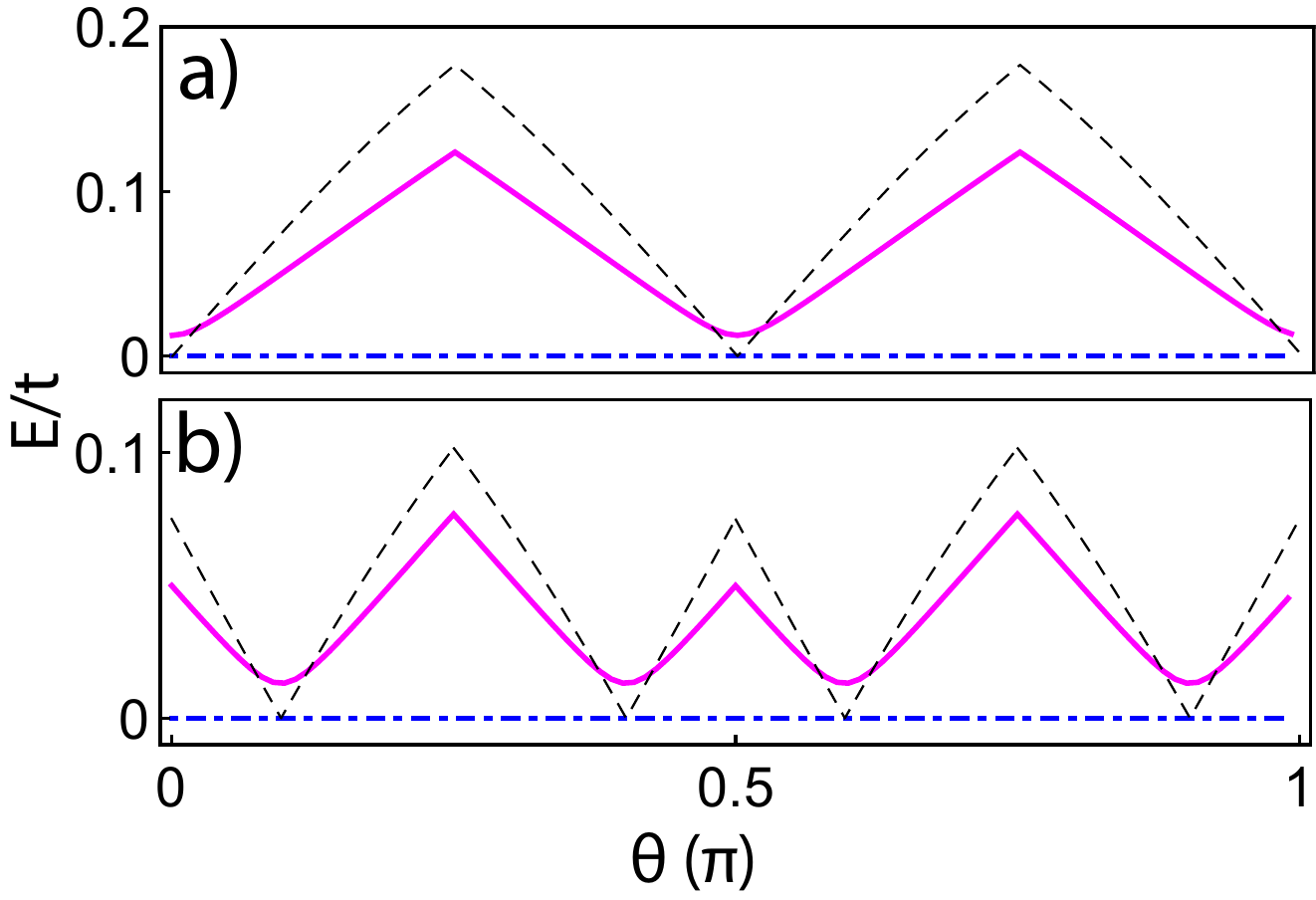}
\caption{Variations of the edge gap with the orientation $\theta$ of the Zeeman field for a $80\times 80$ lattice. Blue (dot dashed) line represents one of the two zero modes, magenta (solid) line reflects the edge gap, and black (dashed) line are edge gaps determined by the minimum of mass terms in Eq.(\ref{eq:Hamiltonian_edge}) for the four edges. $t=1$, $\mu = 3$, $\Delta = 1$, $\phi=0$ and $V = 0.5$. Panel b) An inversion-breaking term $\Delta_s\tau_y$ is added, with $\Delta_s = -0.15$.}\label{fig:fig2}
\end{figure}

\begin{figure*}
\includegraphics[scale=0.45]{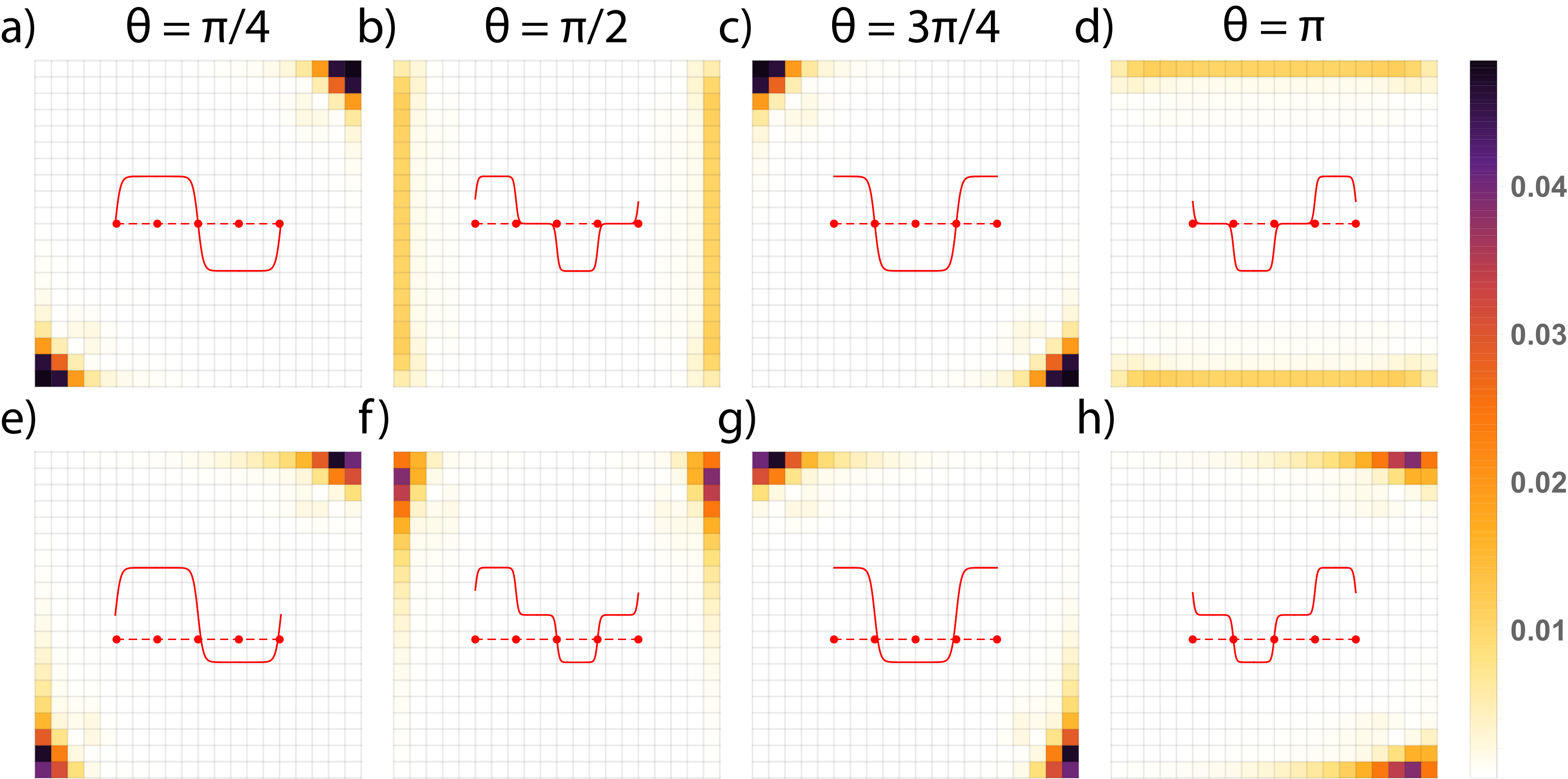}
\caption{Probability density distribution of MZMs for a $20\times 20$ lattice. Insets are the mass profile on 1D system shown in the right panel of Fig.\ref{fig:fig1}, with dotted line representing zero masses and red dots denoting the corners. $t=1$, $\mu = 3$, $\Delta = 1$, $\phi=0$ and $V = 0.5$. For (e)-(h) the term $\Delta\tau_y$ is added, with $\Delta_s = -0.15$.}\label{fig:fig3}
\end{figure*}

So far, we have only considered PBCs, imposed either in one or both directions, corresponding to a system being put on a cylinder or torus. Now we turn to a finite lattice with free boundaries in both directions, in which case the four edges are all well defined. One can expect that the spectrum of edge modes determined by Hamiltonian (\ref{eq:Hamiltonian_edge}) could still be valid at least qualitatively. Indeed, as shown in Fig.\ref{fig:fig2}(a), the edge gap for a finite lattice agrees well with theoretical values on a cylinder geometry--- minimum of the four edge gaps--- given by Eq.(\ref{eq:Hamiltonian_edge}). In addition, we find that two MZMs (only one was shown in Fig. \ref{fig:fig2}) always exist, even when certain edges become gapless, corresponding to $\theta$ being an integer multiple of $\pi\over 2$. A further investigation reveals that the two MZMs are bound at corners on the same diagonal, as can be seen from the probability density plot in Fig.\ref{fig:fig3} (a)-(d). While the Zeeman field rotates in the plane, these MCSs may hop from one corner to anther, and in some special cases reside on certain edges, which is due to the vanishing of mass terms on these edges. When the Zeeman field rotates by $2\pi$, each MCS would complete one revolution around the lattice and return to its original position. To figure out when and where (which corners) these MZMs could form, we need to identify the relative signs of mass terms among the four edges. 


\begin{figure}[b]
\includegraphics[scale=0.38]{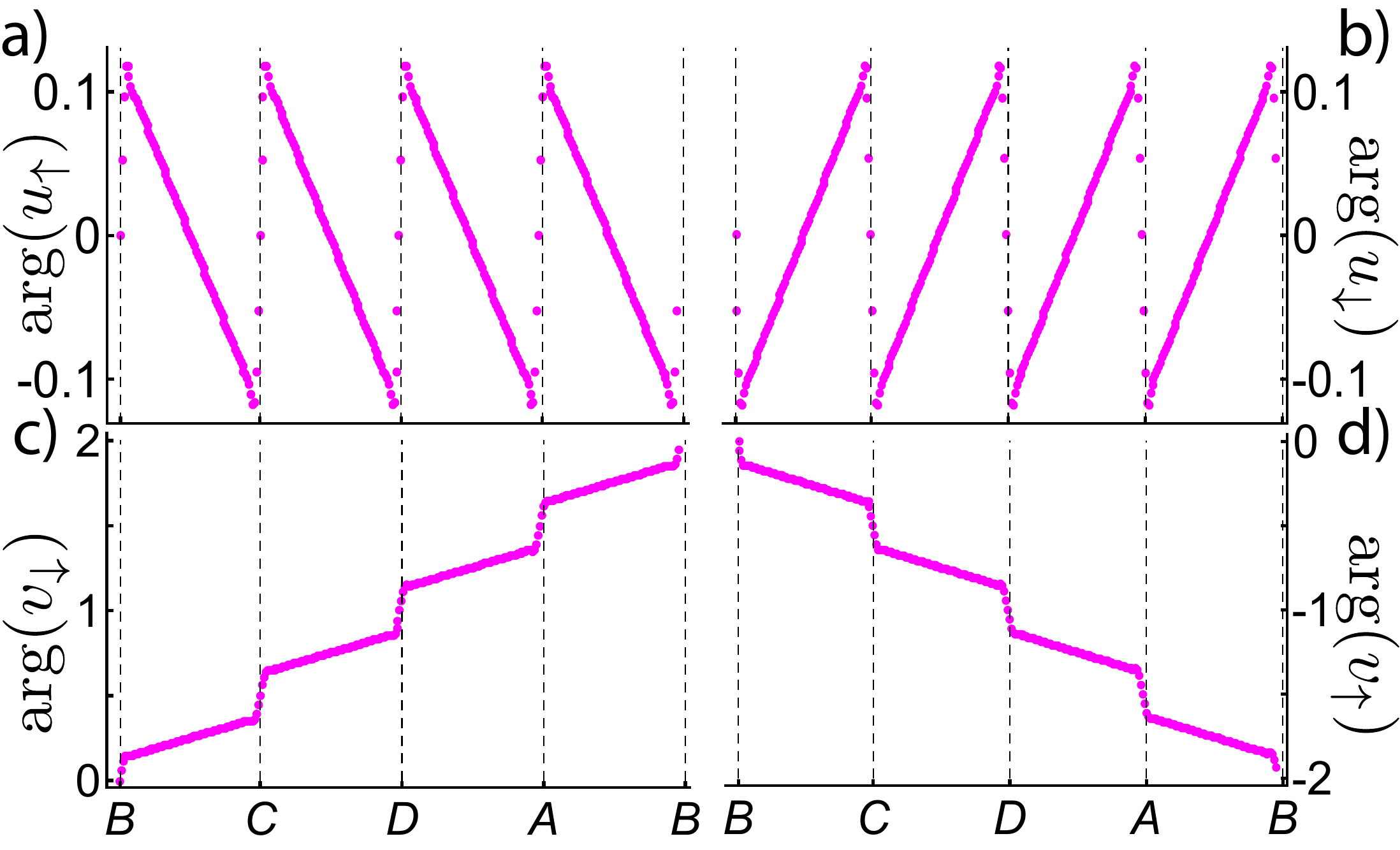}
\caption{Phase for edge modes with lowest positive energy along the boundary sites of a $80\times 80$ lattice. The phase for each component is given by its argument relative to that of site $B$. The values shown on vertical axis are expressed in unit of $\pi$. $t=1$, $\mu = 3$, $\Delta = 1$, $\phi=0$ and $V = 0.5$. }\label{fig:fig4}
\end{figure}

We first note that, without Zeeman fields edge modes in a finite lattice are supposed to circle around all the four edges, with the spinor part on each edge being approximately described by $\psi^{\text{Edge}}_\alpha$ given in Eq.(\ref{eq:wavefunction_edge}), provided the system is large enough. In addition, $C_4$ symmetry defined in Eq.(\ref{eq:c4rotaion}) requires that the spinors for an edge mode flowing on the four edges are connected by unitary transformation $\mathcal U_R$, to be specific,
\begin{equation}
\psi^{\text{BC}}_\alpha= \mathcal U_R^{-1}\psi^{\text{CD}}_\alpha=\mathcal U_R^{-2}\psi^{\text{DA}}_\alpha=\mathcal U_R^{-3}\psi^{\text{AB}}_\alpha.\label{eq:rotation_edge}
\end{equation}
Since $\mathcal U_R^4 = -1$, the spinor part would acquire a minus sign after the mode makes a full circle, in contradiction with single-valuedness of wave functions. Hence there has to be another term in the wave function that also contributes to a minus sign. Notice that for an edge mode with finite energy $E$, the wave function $\Psi^{\text{Edge}}_\alpha$ in Eq.(\ref{eq:wavefunction_edge}) needs to be modified by multiplying an oscillating term $e^{ikr}$ with $k = \frac{2E}{\Delta}$ (the prefactor 2 originates from $1\over 2$ in Hamiltonian (\ref{eq:Hamiltonian_k})). To produce a minus sign, we then expect $kr$ to increase or decrease by an odd multiple of $\pi$ when an edge mode goes back to its original position after a complete revolution, \textit{i.e.}, $kP=(2n+1)\pi$ with $P$ being perimeter of the lattice. Therefore no gapless modes would exist in a finite lattice. This resembles the spinless $p+ip$ TSC on an annulus in Ref.[\onlinecite{Alicea2012}], where edge modes flowing on the inner and outer edges are also gapped, and the infinite-order rotation symmetry of annulus requires the spinor part to rotate all the way while the gapped edge modes flow. In Fig.\ref{fig:fig4}, we plotted the phase evolution on the boundary sites for each component of the two lowest positive edge modes, one for each spin, with wave functions denoted by $\{u_\uparrow, v_\uparrow\}$ and $\{u_\downarrow, v_\downarrow\}$ respectively. As is demonstrated in Fig.\ref{fig:fig4}, the phase experiences an abrupt change around each corner, corresponding to the rotation of spinor part, in agreements with Eq.(\ref{eq:rotation_edge}). Instead, on the edges the phase varies smoothly, which is due to the spatially oscillating term. Clearly, for a mode with positive energy spin-up modes propagate clockwise whereas spin-down modes propagate counterclockwise, both characterized by a constant modulus of wave vector, being $k = \frac{\pi}{P}$, which can be readily evaluated from the slope of phase curves on each edge shown in Fig.\ref{fig:fig4}. 


Now we have established that, i) edge modes for a finite lattice without Zeeman field could be described by $\Psi_\alpha^{\text{Edge}}$ given in Eq.(\ref{eq:wavefunction_edge}), with the spinor part on the four edges being connected to one another through the relation in Eq.(\ref{eq:rotation_edge}), as verified by Fig.\ref{fig:fig4}; ii) edge spectrum could be approximately described by those Hamiltonian obtained on a cylinder geometry shown in Eq.(\ref{eq:Hamiltonian_edge}), as can be seen from Fig. \ref{fig:fig2}. However, since the Hamiltonian for each edge was written in different edge space, we still couldn't determine the relative signs of mass terms. Our strategy is then to transform the basis for each edge space, $i.e.$, $\{\Psi^{\text{Edge}}_{\uparrow}(r),\Psi^{\text{Edge}}_{\downarrow}(r)\}^T$ given in Eq.(\ref{eq:wavefunction_edge}) into the same form. This actually amounts to a rotation of bulk Hamiltonian (\ref{eq:Hamiltonian_k}), with which one could obtain the new edge Hamiltonian following the same procedure while deriving Eq.(\ref{eq:Hamiltonian_edge}). To elaborate this point, let us consider performing a reverse $C_4$ rotation ($C_4^{-1}$) around corner $B$, which would send Edge $CD$ to $CD'$ as shown in Fig. \ref{fig:fig1}. The new bulk Hamiltonian is then $\mathcal H'(\bm k) = \mathcal U_R^{-1}\mathcal H(\mathcal R\bm k)\mathcal U_R$. One can verify that wave functions of MZMs on $CD'$ in absence of Zeeman fields have exactly the same form as in $BC$, that is, the edge space of $CD'$ is also given by $\{\Psi_\uparrow^{\text{BC}},\Psi_\downarrow^{\text{BC}}\}^T$. Then we can project $\mathcal H'({\bm k})$ onto this edge space and obtain the new edge Hamiltonian of $CD'$ by assuming PBC along $CD'$, as we did when obtaining Eq.(\ref{eq:Hamiltonian_edge}). Comparing the mass term in this new edge Hamiltonian with that of Edge $BC$, we could identify the relative signs of masses for them. Similarly, one could perform two or three consecutive $C_4^{-1}$ rotation, and map the other two edges onto the 1D system as well, shown in the right panel of Fig.\ref{fig:fig1}. In this 1D system, there are four sections, each of which corresponds to an edge in the original system, and the wave functions of MZMs on them have the same form. The bulk Hamiltonian relating to each section, however, have different forms, which we summarized as follows,
\begin{eqnarray}
&&BC\rightarrow BC:\ \ \mathcal H'(\bm k) = \mathcal H(\bm k)\label{eq:rotation_bulk}\\
&&CD\rightarrow CD':\ \ \mathcal H'(\bm k) = \mathcal U_R^{-1}\mathcal H(\mathcal R\bm k)\mathcal U_R\nonumber\\
&&DA\rightarrow D'A':\ \ \mathcal H'(\bm k) = \mathcal U_R^{-2}\mathcal H(\mathcal R^{2}\bm k)\mathcal U_R^{2}\nonumber\\
&&AB\rightarrow A'B':\ \ \mathcal H'(\bm k) = \mathcal U_R^{-3}\mathcal H(\mathcal R^3\bm k)\mathcal U_R^3.\nonumber
\end{eqnarray}
By projecting all the four bulk Hamiltonian $\mathcal H'(\bm k)$ listed in Eq.(\ref{eq:rotation_bulk}) onto this basis $\{\Psi_\uparrow^{\text{BC}},\Psi_\downarrow^{\text{BC}}\}$, we could obtain the corresponding edge Hamiltonian for each section on the 1D system, which takes the following concise form
\begin{equation}
\mathcal H^{\text{Edge}}_{\text{1D}} = -\Delta k_x\eta_z - V\sin(\theta+\phi+2\varphi_n)\eta_y,\label{eq:Hamiltonian_edge1}
\end{equation}
Eq.(\ref{eq:Hamiltonian_edge1}) now describes a massless spinor field in 1D space subject to a spatially varying mass field as schematically illustrated in the right panel of Fig. \ref{fig:fig1}.  Alternatively, one could also follow the method introduced in Ref.[\onlinecite{Zhang2013a}] and [\onlinecite{Song2017}] to identify the mass terms on each edge, as we did in Appendix \ref{sec:appendix}. Essentially, this method also transforms the wave functions of MZMs for different edges into the same form, but through a rotation of basis $\Psi_{\bm k}$ of bulk Hamiltonian (\ref{eq:Hamiltonian_k}) (both in real space and spinor space). This could be seen as a \emph{passive} point of view to look at the $C_4^{-1}$ rotation performed on the system, in contrast to the \emph{active} point of view we took while deriving Eq.(\ref{eq:rotation_bulk}). Clearly these two methods lead to the same results.

From Eq.(\ref{eq:Hamiltonian_edge1}) we can immediately read off the masses for each edge, $i.e.$, $m_{\text{Edge}} = - V\sin(\theta+\phi+2\varphi_n)$, where $\varphi_n = \frac{n}{4}\pi$ and $n$ corresponds to edge indices being the same as in Eq.(\ref{eq:wavefunction_edge}). For two opposite edges in the square lattice, $\varphi_n$ differs by $\pi/2$, and hence the mass sign reverses. Specifically, we would have $\text{sgn}(m_{\text{AB}})\text{sgn}(m_{\text{CD}})=\text{sgn}(m_{\text{BC}})\text{sgn}(m_{\text{DA}})=-1$. Therefore, if all the four edges are gapped out by the field, one can verify that there would always be two corners where adjacent edges with opposite masses intersect. That at which two corners the mass may change sign depends on the specific value of system parameters $\theta+\phi$. Suppose, for instance $\theta+\phi=\pi/4$, we would have, $\text{sgn}(m_{\text{BC}})=\text{sgn}(m_{\text{CD}})=-\text{sgn}(m_{\text{DA}})=-\text{sgn}(m_{\text{AB}})=1$. Clearly, mass sign reverses at point $B$ and $D$. When such a mass kink forms, a single MZM protected by particle-hole symmetry would emerge, as have been well established in the seminal paper by Jackiw and Rebbi\cite{Jackiw1976}. In the insets of Fig.\ref{fig:fig3}, we plotted the mass profile given by Eq.(\ref{eq:Hamiltonian_edge1}) for the 1D system, which apparently demonstrates that, whenever a kink forms at certain corner there would be one single MZM localized around. The decaying length for these MCSs along the two intersecting edges is inversely proportional to the mass of each edge. Since the mass field is a sine function of the orientation $\theta$ of the in-plane Zeeman field, kinks and MCSs accompanied with them could thus be driven across edges and hop from one corner to a neighboring one, as we have already seen in Fig. \ref{fig:fig3}(a)-(d). After the field rotates by $2\pi$, the mass field reverts to its original configuration, during which MCSs exactly make a full circle along the boundary.

The fact that mass terms for opposite edges always manifest reverse signs is because two consecutive $C_4^{-1}$ rotation or a $C_2^{-1}$ rotation of bulk Hamiltonian, like the one in $DA\rightarrow D'A'$ from Eq.(\ref{eq:rotation_bulk}), sends $(\sigma_x,\sigma_y)$ to $(-\sigma_x,-\sigma_y)$, thus being equivalent to reversing the orientation of in-plane Zeeman fields. In other words, the Zeeman-field induced mass terms are odd under $C_2$ rotation. Moreover, the two corners that support MZMs reside on the same diagonal of the lattice, as shown in Fig.\ref{fig:fig3} (a) and (c). Also note that, $\sigma_x$ and $\sigma_y$ term both preserve the inversion symmetry $\mathcal I$ (differing with $C_2$ rotation by a unitary transformation $\tau_z\sigma_z$) as we have established in Eq.(\ref{eq:inversion}). Consequently, the distribution of MZMs also preserves inversion symmetry as shown in Fig.\ref{fig:fig3} (a)-(d). 

Interestingly, one may verify that $\sigma_x$, $\sigma_y$ and $\tau_y$ are the only three $\bm k$-independent terms that could gap out the edges of a $p \pm ip$ superconductor. Contrary to $\sigma_x$ and $\sigma_y$ term, a perturbation like $\Delta_s\tau_y$ is even under $C_2$ rotation and meanwhile breaks inversion symmetry $\mathcal I$. We can expect it to introduce a mass term being the same on opposite edges. Indeed, upon projecting this $\tau_y$ term onto edge space, we find it gives rise to an additional mass term on each edge Hamiltonian, given by $-\Delta_s\eta_y$, independent of the orientation of edges. Again, we compare this mass term derived on a cylinder geometry with numerical results obtained for a finite lattice, which agrees well as shown in Fig.\ref{fig:fig2}(b). Since $\tau_y$ term only leads to a global shift of mass profile in Hamiltonian (\ref{eq:Hamiltonian_edge1}), kinks in the mass profile could then survive provided the perturbation is weak enough comparing to the Zeeman field. As a result, MCSs continue to exist as can be found in Fig. \ref{fig:fig2}(b) and Fig. \ref{fig:fig3}(e)-(h). However, the distribution of MZMs no longer preserves inversion symmetry $\mathcal I$. Increasing $\tau_y$ term further would eliminate the kinks and drives the system into a trivial phase.

So we have seen a transition from a traditional or first-order TSC to a second-order one. Now we shall briefly comment on the distinctions between them, and justify the name of \emph{second order}. According to the tenfold classification\cite{Schnyder2008,Chiu2016}, a $p\pm ip$ TSC in 2D belongs to DIII class and is characterized by a $\mathbb Z_2$ index, which counts the parity of Majorana Kramers pairs propagating along each edge. An in-plane Zeeman field breaks TRS and as a result the system falls into D class, characterized by a $\mathbb Z$ index. Since we start from a $p\pm ip$ TSC residing in nontrivial phases, a weak in-plane field would gap out the Majorana pairs. This can be seen from the effective edge Hamiltonian (\ref{eq:Hamiltonian_edge}) obtained on a cylinder geometry, and it is obvious that there are no MZMs propagating on the edges. Naively, this should correspond to a trivial phase in the tenfold classification scheme. However, we have demonstrated that it's actually not trivial, and MZMs could emerge at corners instead, which suggests the system cannot be smoothly connected to the vacuum, and thus it's still in the nontrivial phase. In contrast to first-order topological phases where gapless modes have to emerge at boundaries (or at least along certain directions for weak topological phases), a second-order phase have gapped boundaries whereas topologically protected gapless modes occur at the intersection of topologically different boundaries, that is, at the \emph{boundary of boundaries}, hence the name \emph{second order}. In this article, the topology of boundaries are manifested through signs of masses for each edge. It would be very interesting to characterize such a second-order phase by a bulk invariant, especially when all the symmetries except for particle-hole symmetry are broken, and we leave this for future investigations.

\begin{figure}[h]
\includegraphics[scale=0.45]{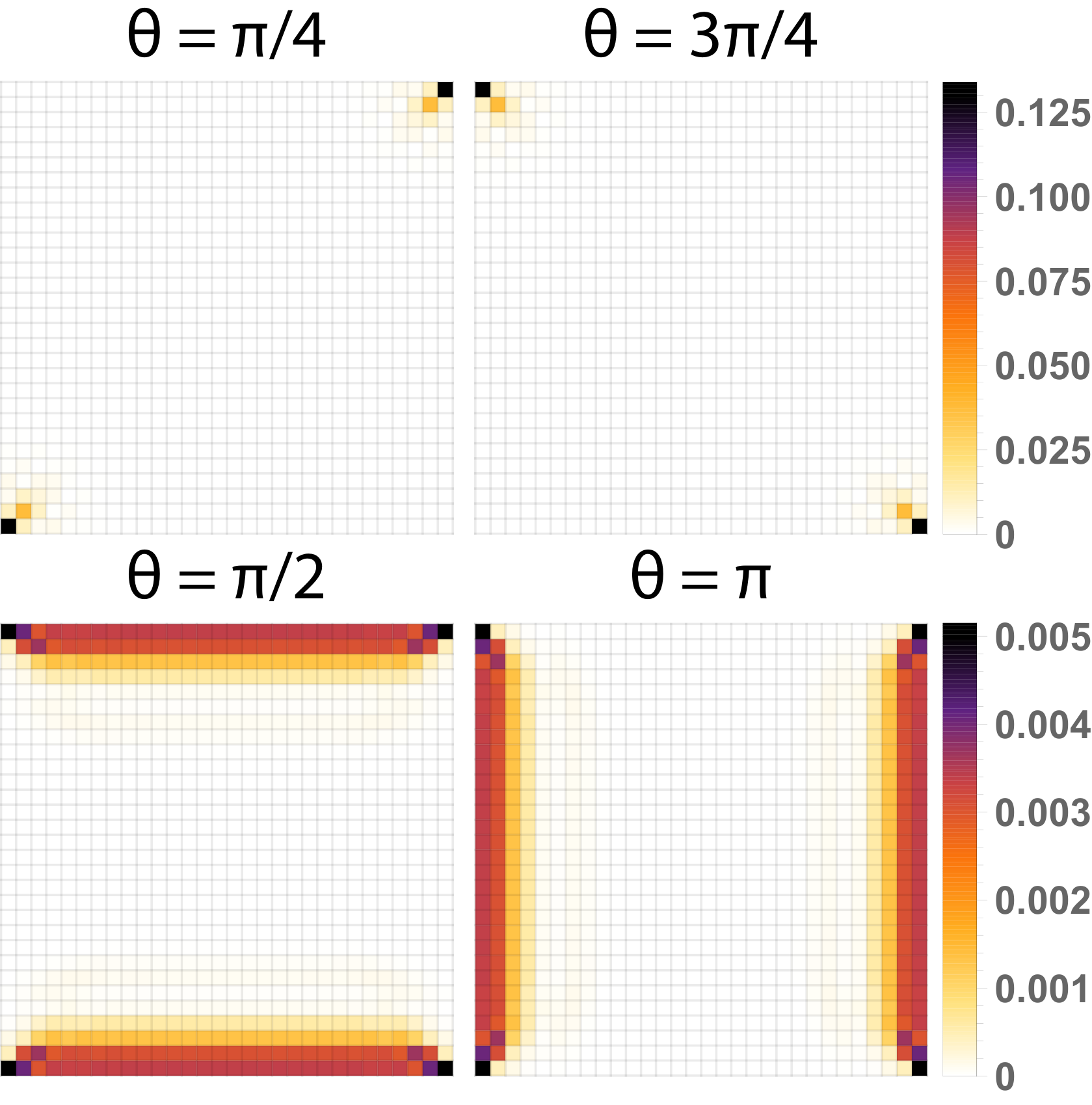}
\caption{Probability density plot of MZMs for a $30\times 30$ lattice introduced in Ref.[\onlinecite{Zhang2013b}] with an additional in-plane magnetic field applied. Parameters in the lattice Hamiltonian therein are chosen to be, $t=1$, $\lambda_R=1$, $\mu = -1$, $\Delta_0= -\Delta_1 = -2$. Strength of the Zeeman field $V = 1$.}\label{fig:fig5}
\end{figure}

\section{Discussion and conclusion} 

A $p\pm ip$ superconductor is the simplest model of time-reversal-invariant TSCs in 2D. In Ref.[\onlinecite{Zhang2013b}], Zhang \textit{et al.} introduced a Rashba semiconductor/nodeless iron-based superconductor heterostructure as a promising platform to realize a 2D TSC protected by TRS. Similar to the $p\pm ip$ superconductor, we consider applying an in-plane Zeeman field $\bm V = (V\cos\theta,V\sin\theta,0)$ to this system, and find that MZMs could also occur at corners, as is shown in Fig. \ref{fig:fig5}. Furthermore, just like in a $p\pm ip$ superconductor, rotating the magnetic field in the plane would also move MCSs among the four corners.

Other promising candidates for time-reversal-invariant TSC in 2D include Rashba bilayer with interlayer interactions\cite{Nakosai2012}, Rashba semiconductor with two bands differing by a $\pi$-phase shift in $s$-wave superconducting order parameter\cite{ShuDeng2012}, and some other interesting systems\cite{Wang2014,Yang2015,Midtgaard2017}. The low energy theory for them could all be described by a $p\pm ip$ superconductor. An in-plane magnetic field applied in these systems, however, may take distinct forms when projected to the low energy theory of edges, and thus may or may not gap out them. It would be interesting to inspect these specific systems and see if an in-plane magnetic field could give rise to MCSs. 

In conclusion, we demonstrate that a 2D $p\pm ip$ superconductor protected by TRS could support MCSs when an in-plane magnetic field is applied. By mapping all the four edges onto a 1D system, we identified the relative signs of Zeeman-field-induced masses among the four edges. MZMs form at the intersection of two adjacent edges when the mass flips sign between the two edges. Two MCSs that separate from each other in space could be induced in this model, and the positions of them could be tuned simply by rotating the in-plane magnetic field. When the field rotates by $2\pi$, each MCS confined on the boundary would make a full circle around the system center accordingly. The simplicity manifested while tuning MCSs makes the system a potential platform to perform braidings of MZMs\cite{Nayak2008}.

\textit{Note added.}  Recently we became aware of related work in Refs.[\onlinecite{Khalaf2018},\onlinecite{Geier2018}].

\acknowledgements
I would like to thank W. Chen for many stimulating discussions. This work was supported by National Science Foundation of China under Grant No. 11704305.\\

\appendix
\section{Mass of an edge along an arbitrary direction}\label{sec:appendix}

\begin{figure}[b]
\includegraphics[scale=0.35]{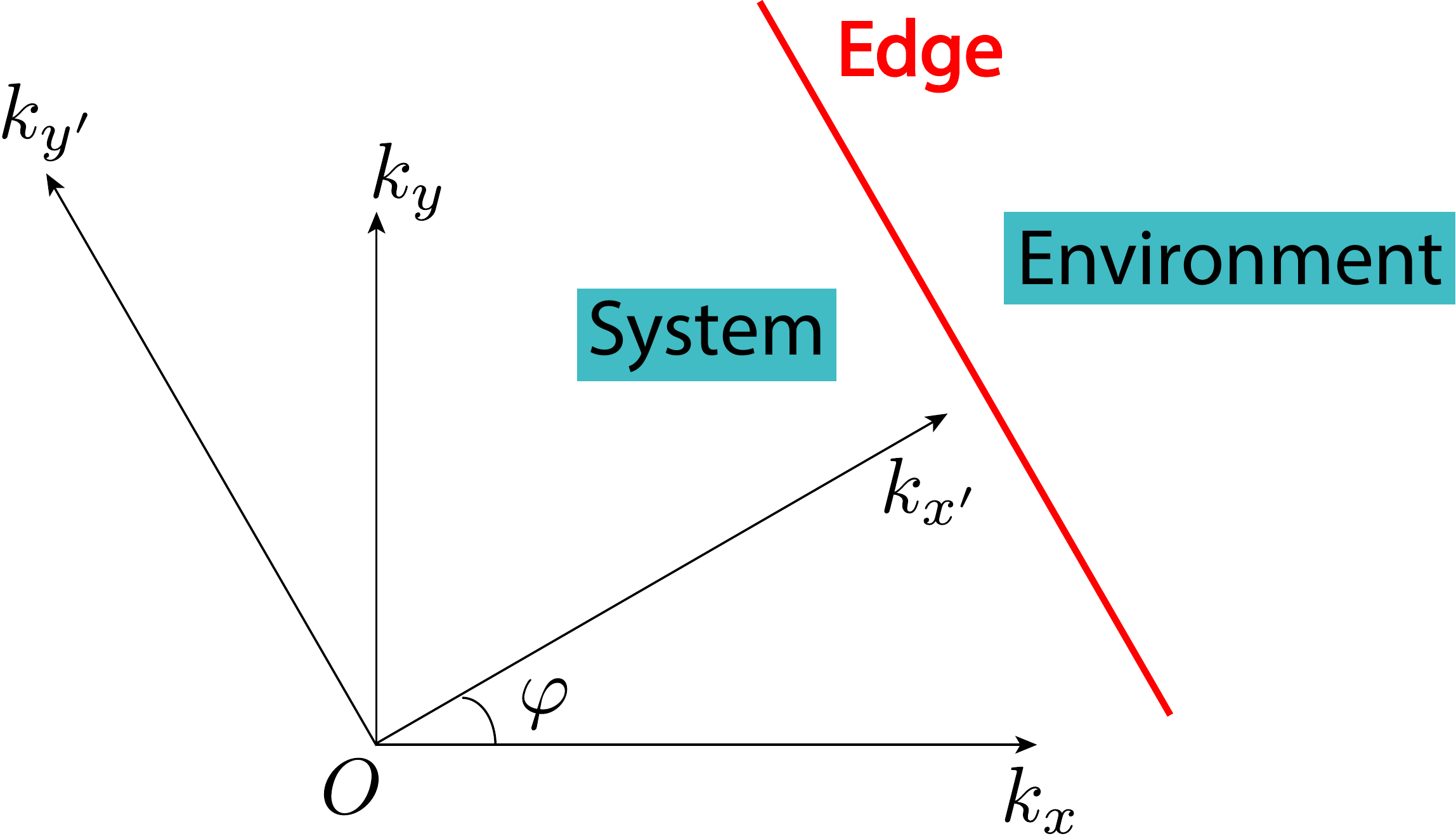}
\caption{An arbitray edge of the system considered in the main text. Two sets of Cartesian coordinate systems, $O$-$k_xk_y$ and $O$-$k_{x'}k_{y'}$ are shown, related by an in-plane rotation of $\varphi$. Axis-$k_{y'}$ is along the direction of the edge.}\label{fig:fig6}
\end{figure}

In this appendix, we will provide an alternative method to derive the Dirac mass on each edge, following the procedure in Ref.[\onlinecite{Zhang2013a}] and [\onlinecite{Song2017}]. Let us consider an edge along an arbitrary direction as shown in Fig. \ref{fig:fig6}. Denote a point in reciprocal space to be $(k_x,k_y)$ and $(k_{x'},k_{y'})$ in the two coordinate systems, $O$-$k_xk_y$ and $O$-$k_{x'}k_{y'}$ respectively. Then we have
\begin{equation}
k_x = k_{x'}\cos \varphi - k_{y'}\sin\varphi, k_y = k_{x'}\sin\varphi + k_{y'}\cos\varphi.\label{eq:rotation}
\end{equation}
Suppose the bulk Hamiltonian (\ref{eq:Hamiltonian_k}) is written in $O$-$k_xk_y$ system. In the following, we shall rewrite the bulk Hamiltonian (\ref{eq:Hamiltonian_k}) in $O$-$k_{x'}k_{y'}$ system, and derive the edge Hamiltonian in this system. First, we substitute Eq.(\ref{eq:rotation}) into the bulk Hamiltonian (\ref{eq:Hamiltonian_k}) and linearize the resulting Hamiltonian around $\Gamma$ point, which leads to
\begin{eqnarray}
H &=& \frac{1}{2}\sum\limits_{\bm k'} \Psi_{\bm k'}^\dagger \mathcal{H}_0(\bm k')\Psi_{\bm k'}\label{eq:H0}\\
\mathcal H_0(\bm k') &=& (\mu-4t)\tau_z - \Delta\tau_x[(k_{x'}\cos\varphi-k_{y'}\sin\varphi)\sigma_x\nonumber\\
&-&(k_{x'}\sin\varphi+k_{y'}\cos\varphi)\sigma_y]+\bm{\tilde V\cdot \sigma}\nonumber
\end{eqnarray}
Applying a further rotation to the basis $\Psi_{\bm k'}$ in spinor space, that is, $\tilde\Psi_{\bm k'}= \mathcal U\Psi_{\bm k'}$, with $\mathcal U = e^{-i\frac{\varphi}{2}\sigma_z}$, the Hamiltonian could be transformed to a similar form as in Eq.(\ref{eq:Hamiltonian_k}) given by
\begin{eqnarray}
H &=& \frac{1}{2}\sum\limits_{\bm k'} \tilde\Psi_{\bm k'}^{\dagger} \mathcal{H}_1(\bm k')\tilde\Psi_{\bm k'}\\
\mathcal H_1(\bm k') &=& \mathcal U \mathcal H_1(\bm k')\mathcal U^{-1}\label{eq:Hamiltonian_k1}\nonumber\\
&=&(\mu-4t)\tau_z - \Delta\tau_x(\sin k_{x'}\sigma_x-\sin k_{y'}\sigma_y) + \tilde{\bm V_1}\cdot\bm\sigma,\nonumber
\end{eqnarray}
where the effective Zeeman field $\tilde{\bm V_1} = (V\cos(\theta+\phi+\varphi),V\sin(\theta+\phi+\varphi),0)$, with $\theta$ and $\phi$ being the same as in Eq.(\ref{eq:Hamiltonian_k}). Assume chemical potential $\mu$ varies slowly across the edge, with $\mu$ less (greater) than $4t$ inside (outside) the system. Imposing PBC along the edge (axis-$k_{y'}$) and OBC perpendicular to the edge (axis-$k_{x'}$), we could replace $k_{x'}$ in Eq.(\ref{eq:Hamiltonian_k1}) by $-i\partial_{x'}$. In absence of Zeeman fields, it's straightforward to obtain the wave functions of gapless edge modes localized on the edge as we did in Sec.\ref{sec:Edge}. Then we could project the Hamiltonian in Eq.(\ref{eq:Hamiltonian_k1}) to the edge space spanned by the gapless modes, which leads us to the effective edge Hamiltonian for an edge pointing along an arbitrary direction,
\begin{equation}
\mathcal H_1^{\text{Edge}}(k_{y'}) = -\Delta k_{y'}\eta_z + V\cos(\theta+\phi+\varphi)\eta_y.\label{eq:edge_1}
\end{equation}
From Eq.(\ref{eq:edge_1}) we can immediately read off the mass for an edge, $i.e.$, $m_{\text{Edge}} = V\cos(\theta+\phi+\varphi)$, which clearly depends on the orientation of it. Note that $\varphi=0$ corresponds to Edge $CD$ in Fig.\ref{fig:fig1} with $\varphi_n = \frac{3}{4}\pi$ in Eq.(\ref{eq:Hamiltonian_edge1}). One may verify that the mass term given in Eq.(\ref{eq:edge_1}) is consistent with that in Eq.(\ref{eq:Hamiltonian_edge1}).

\bibliography{mcs.bib}

\end{document}